\documentclass[prb,showpacs,twocolumn]{revtex4}
\usepackage{array,amsmath,amssymb,graphicx}
\pdfoutput=1

\begin{document}



\title{Kinetics of Exciton Emission Patterns and Carrier Transport}

\author{Sen Yang$^1$, L. V. Butov$^1$, L. S. Levitov$^2$, B. D. Simons$^3$, A. C. Gossard$^4$}
\affiliation{$^1$Department of Physics, University of California at San
Diego, La Jolla, CA 92093-0319, USA}
\affiliation{$^2$Department of Physics, Massachusetts Institute of
Technology, Cambridge, Massachusetts 02139, USA}
\affiliation{$^3$Cavendish Laboratory, Madingley Road, Cambridge CB3
OHE, United Kingdom}
\affiliation{$^4$Materials Department, University of California at Santa
Barbara, Santa Barbara, CA 93106-5050, USA}

\date{\today}

\begin{abstract}
We report on the measurements of the kinetics of expanding and collapsing rings in the exciton emission pattern. The rings are found to preserve their integrity during expansion and collapse, indicating that the observed kinetics is controlled by charge carrier transport rather than by a much faster process of exciton production and decay. The relation between ring kinetics and carrier transport, revealed by our experiment and confirmed by comparison with a theoretical model, is used to determine electron and hole transport characteristics in a contactless fashion.
\end{abstract}


\maketitle

Transport characteristics of carriers are central to the properties of many materials. Conventional approaches, which are based on electrical measurements, provide a wealth of information on transport coefficients. However, since these methods rely on contacts for injecting current and measuring voltage, they can become problematic in some systems, for instance when electrical conductivity is low or when good contacts are difficult to make. In such cases optical methods take the central stage as a tool for investigating the carrier transport. Optical methods are contactless and can be successfully applied even to systems with low carrier density. In addition, since these methods do not rely on the carrier charge, they can be employed to study transport of neutral particles. A variety of optical methods, including transient grating, time-of-flight, pump-probe, and imaging techniques, were developed to probe transport of electrons, holes, and excitons (see Ref.\cite{Smith89} and references therein).

Optical methods can be useful for studying electron and hole transport in coupled electron-electron and electron-hole layers. Making separate contacts to the layers for transport measurements becomes challenging for systems with layer separation under several tens of nm. Yet, this is the most interesting case, which is characterized by high interlayer correlations responsible for the formation of novel electronic states \cite{Sivan92, Eisenstein92, Lay94, MacDonald01, Pellegrini07, Tiemann08, Croxall08, Seamons09}. In particular, small layer separation is required for the radius $a_B$ of an indirect exciton formed from an electron and a hole in separated layers to be small and its binding energy $E_X$ to be large, leading to a high critical temperature for the exciton condensation $T_0$: For the BCS-like exciton condensate, $T_0$ scales with $E_X$ \cite{Keldysh65}; for the BEC-like exciton condensate, maximum $T_0$ is achieved at densities $n_X\sim 1/a_B^d$, increasing upon reducing $a_B$, where $d$ is the dimensionality of the system \cite{Keldysh68}.

In this paper, we demonstrate that kinetics of the exciton emission patterns in quantum well structures provides a method for probing carrier transport, essentially for any layer separation. The patterns include the inner rings, external rings, localized bright spots (LBS), and macroscopically ordered exciton state \cite{Butov02}. The kinetics of the inner ring provides insight into exciton transport \cite{Ivanov06}, whereas the kinetics of the external rings and LBS rings, on which we focus here, gives information on electron and hole transport \cite{compare}.

The use of ring kinetics as a vehicle for studying carrier transport is made possible by the observation that ring expansion and collapse occur on microsecond time scales which are much longer than the ring formation time. Our observation that exciton rings preserve their integrity during expansion and collapse is a strong indication that the characteristic times for the latter are much slower than those for self-organization of electrons, holes and excitons into ring-like patterns. This crucial property, which was assumed in models of ring formation \cite{Butov04,Rapaport04,Haque06}, has not been verified experimentally prior to present work. Our observation is consistent with the previous work \cite{Chen05} which concluded, without measuring ring kinetics, that the response of the external ring to the modulation of the applied voltage and excitation power is slower than exciton recombination.

\begin{figure*}\includegraphics[width=1\textwidth]{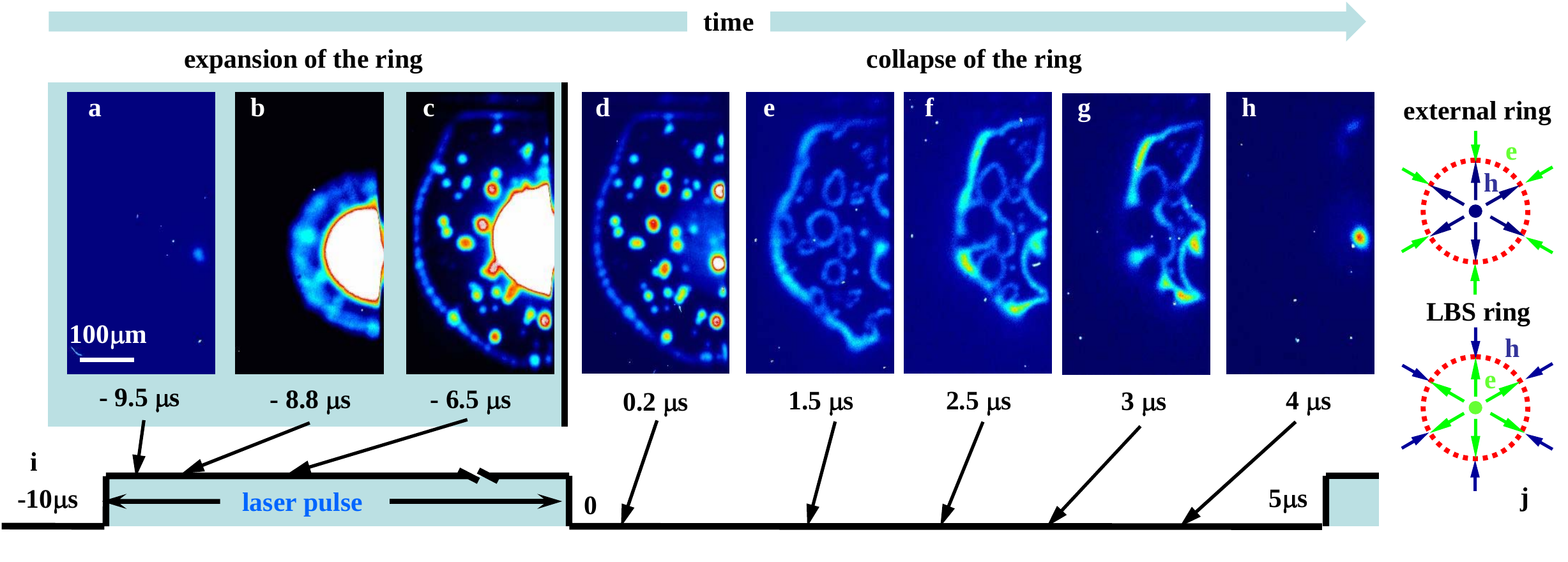}
\caption{(color online) (a-h) Kinetics of exciton pattern formation. Emission images for indirect excitons at different time delays obtained with time-integration window of 200 ns. Zero time is chosen at the end of the laser pulse. Gate voltage $V_g$ =1.235 V, laser peak power $P_{ex}=1.82$ mW. (i) Schematic of the laser pulse duty cycle sequence. (j) Schematic of carrier transport in the external ring and LBS ring (see text).} \label{1}
\vspace{-4mm}
\end{figure*}

The coupled quantum well structure (CQW) used in these experiments was grown by molecular beam epitaxy. It contains two 8\,nm GaAs QWs separated by a 4\,nm Al$_{0.33}$Ga$_{0.67}$As barrier and surrounded by 200\,nm Al$_{0.33}$Ga$_{0.67}$As layers (for details on the CQW see \cite{Butov02}). The recombination lifetime of the indirect excitons in the CQW is about 50\,ns \cite{Hammack07}. The measurements were performed using time-resolved imaging with 200 ns integration window and $2 \mu$m spatial resolution at $T = 1.4$\,K. The electrons and holes were photogenerated using rectangular excitation pulses of a semiconductor laser at 1.95\,eV, above the Al$_{0.33}$Ga$_{0.67}$As gap ($\sim 1.93$eV), with a $8 \mu$m spot. The pulse width was $10 \mu$s, with the edge sharpness better than 1\,ns, and the repetition frequency of 67 KHz (Fig. 1i). The period and duty cycle were chosen to provide time for the pattern to approach equilibrium during the laser pulse and to allow complete decay of the emission between the pulses. Different pulse widths and repetition frequencies yield similar results. Time-dependent emission images were acquired by a nitrogen-cooled CCD camera after passing through a time-gated PicoStar HR TauTec intensifier with a time-integration window of 200\,ns. An $800 \pm 5$ nm interference filter, chosen to match the indirect exciton energy, was used to remove the low-energy bulk emission and high-energy direct exciton emission. The spectral filtering and time-gated imaging provided the direct visualization of the spatial intensity profile of the indirect exciton emission as a function of the delay time $t$ after the laser pulse.

Figure 1a shows typical time evolution of the exciton emission pattern. After the start of a rectangular excitation pulse, the exciton ring expands (Fig. 1a-c) approaching a steady state; simultaneously, several
LBS rings appear inside the expanding ring (cf. \cite{Butov04}). After the excitation pulse ends, the external ring collapses while the LBS rings expand (Fig. 1d-h). The time evolution of the external ring radius for different values of excitation power $P_{ex}$ and gate voltage $V_g$ is shown in Fig. 2a,b. To compensate for the deviation from perfect circular shape, the ring radius was estimated from the net area enclosed by the ring. The time evolution of the LBS ring radius for different LBS and $V_g$ is shown in Fig. 3c,d.

\begin{figure*}\includegraphics[width=1\textwidth]{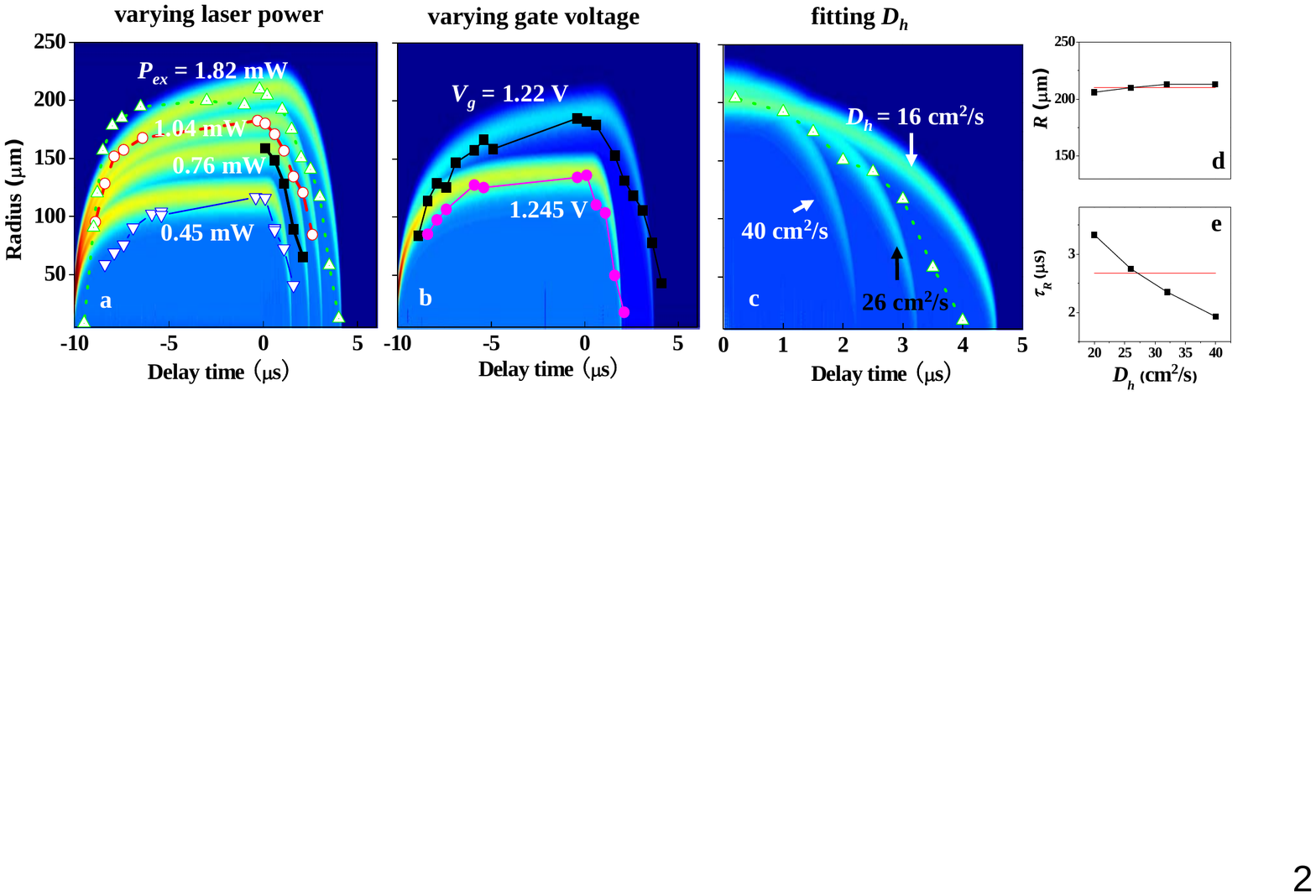}
\caption{(color online) Expansion and collapse of the exciton ring. (a) Measured kinetics of the ring radius for several values of excitation power $P_{ex}=0.45, 0.76, 1.04,$ and 1.82 mW at $V_g=1.235$ V (triangles, squares, circles), shown together with simulated kinetics, Eq.(\ref{eq1}). Results of four different simulations are combined together, for hole source values $I_p = 17.5, 26.25, 32.5,$ and 42.5 (from bottom to top) and $\gamma = 4.2 \times 10^6$ s$^{-1}$ [for definition of units see APPENDIX \ref{app:unit}]. (b) Measured kinetics of the external ring radius for $V_g=1.22$ and 1.245 V at $P_{ex}=0.76$ mW (circles and squares) and simulated kinetics for different electron sources $\gamma = 2.85$ (larger radius) and $5.4 \times 10^6$ s$^{-1}$ (smaller radius) at $I_p=26.25$. The values of other parameters used in the simulations in (a,b) were kept fixed: $D_e=80$ cm$^2$/s, $D_h=26$ cm$^2$/s, $n_{b}=1$, and $w=4$. (c) Simulated kinetics for $D_h=16, 26,$ and 40 cm$^2$/s and the measured kinetics for $P_{ex}=1.82$ mW with the rest of the parameters the same as in (a). The simulations are shifted so that $t=0$ corresponds to the time when the ring radius is maximum. The value $D_h=26$ cm$^2$/s gives the best fit to the measured kinetics. (d,e) Ring radius $R$ and collapse time $\tau_R$ in simulations vs. $D_h$ with the rest of the parameters the same as in (c) (squares). Red line marks the experimental result in (c).} \label{2}
\vspace{-4mm}
\end{figure*}

The observed kinetics was simulated using the model based on the in-plane spatial separation of electrons and holes \cite{Butov04,Rapaport04,Haque06}. The mechanism for the external ring formation can be summarized as follows. The electrons and holes, photogenerated by the photons with the energy above the bandgap of the Al$_{0.33}$Ga$_{0.67}$As layers, can travel in the direction perpendicular to the CQW plane, with some of the carriers being trapped in the CQW. Because the holes are heavier than electrons, the hole trapping to the CQW is more efficient, and therefore an imbalance between holes and electrons in the CQW is created under such photoexcitation. The photoexcited electrons recombine with an equal number of photoexcited holes, producing emission in the vicinity of the excitation spot. At low temperatures and densities, those electrons and holes can bind to form excitons, which can travel away from the excitation spot and cool down to the lattice temperature resulting in the formation of the inner ring in the emission pattern \cite{Butov02,Ivanov06}. The remaining holes, which were photocreated in CQW in excess of the photocreated electrons, diffuse away from the laser excitation spot in the CQW plane, as illustrated in Fig.1j \cite{Butov04}. The diffusing holes recombine with the ambient electrons present in the entire CQW plane due to an electric current through the structure. This process depletes electrons in the vicinity of the laser spot, creating a hole-rich region. At the same time, a spatial nonuniformity in the electron distribution accumulates, causing a counterflow of electrons towards the laser spot. Excitons created at the interface between the inner hole-rich region and the outer electron-rich region give rise to a ring-like emission pattern, the external ring.

This mechanism of ring formation can be modeled by a system of coupled equations:
\begin{equation}\label{eq1}
   \begin{array}{l}
   \partial_t n=D_n\nabla^2n-wnp+\gamma(n_b-n),  \\
   \partial_t p=D_p\nabla^2p-wnp+I_{p}\delta(r),  \\
   \end{array}
\end{equation}
with $n$ and $p$ the electron and hole densities, $D_e$ and $D_h$ the diffusion coefficients, and $w$ the rate of electron and hole binding to form an exciton. Hole production at the excitation spot is described by the term $J_h = I_p \delta(r)$. Electron production occurs over the entire plane; the source $J_e = \gamma (n_b - n)$ is given by the difference of the currents in and out of the CQW, with $\gamma (r)$ the electron escape rate and $n_b(r)$ the background electron density in the CQW in the absence of photoexcitation. A stationary solution of Eq. (\ref{eq1}), with spatially independent $n_b$ and $\gamma$, exhibits electron-rich and hole-rich regions separated by a sharp interface where the exciton density $n_X \propto np$ is peaked, corresponding to the external ring \cite{Butov04}(APPENDIX \ref{app:accuracy}).

Here, we employ the model (\ref{eq1}) to study the ring kinetics. The solutions of Eq.(1) were used to obtain exciton concentration as $n_X \propto np$. The time-dependent profile of $n_X$, presented in Fig. 2 along with the experimental data, reproduces all essential features of the observed kinetics: ring expansion, first rapid, then more slow, followed by collapse when the laser source is turned off. The data obtained for different conditions, such as $P_{ex}$ and $V_g$ in Fig. 2a,b, were fitted using one set of parameters $D_h$, $D_e$, $w$, and $n_b$.

The expansion and collapse of the ring occurs on relatively long time scales, in the range of microseconds, controlled by the carrier diffusion in the sample. This is much longer than the tens of nanoseconds estimated for the inner ring, where decay is limited by the exciton lifetime \cite{Hammack07}. This sensitivity to diffusion makes the external ring an effective probe of carrier transport.

An increase in the amount of holes created by a higher laser excitation results in an increase of both the ring radius and the collapse time (Fig. 2a). In contrast, increasing gate voltage results in an increase in the number of ambient electrons, which reduces both the ring radius and the collapse time (Fig. 2b). These dependences, observed in experiment, are borne out by the simulations.

\begin{figure*}\includegraphics[width=1\textwidth]{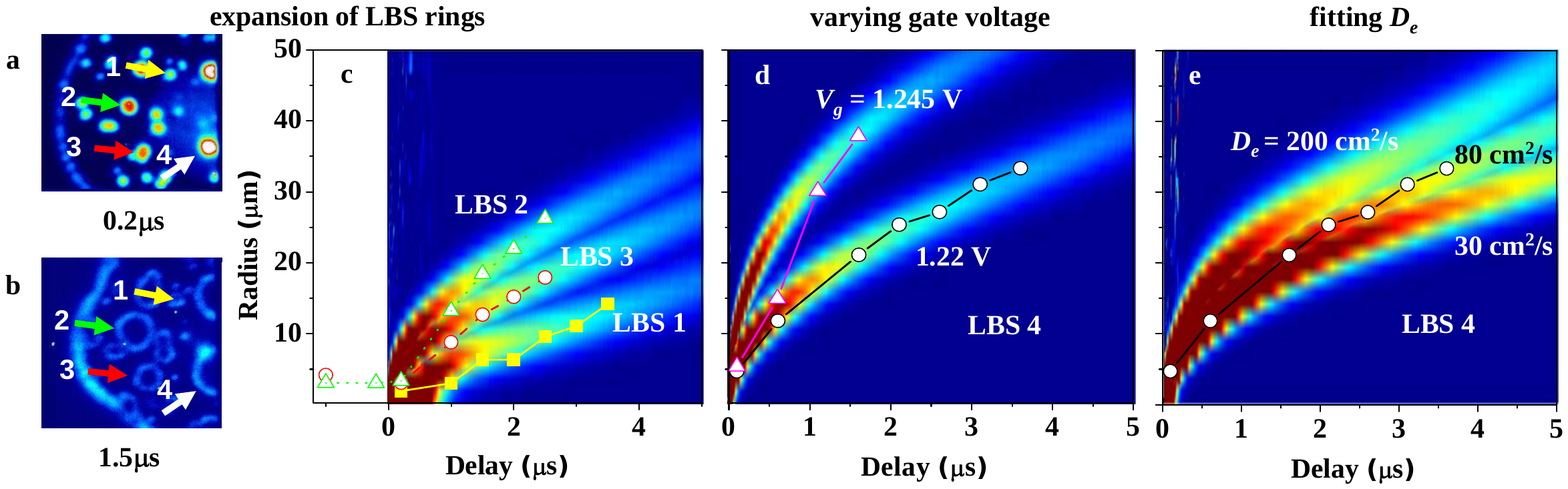}
\caption{(color online) Kinetics of LBS rings. (a,b) Emission images for indirect excitons at time delays $0.2$ and 1.5 $\mu s$ after the laser pulse end. The arrows mark four LBS rings. (c) Measured kinetics of the ring radius for LBS 1, LBS 2 and LBS 3 [yellow, green and red arrows
in (a,b)] at $V_g = 1.235$ V and $P_{ex} = 1.82$ mW (triangles, circles, squares). Note that the brighter LBS in (a) expands faster. Simulated kinetics for different electron source values at the LBS center $n_{bc}=4000, 7000,$ and 8400 (from bottom to top), $n_{bo}=0.05$ away from the LBS center, and the hole density at the pulse end $p_b=10$ (colorscale curves). (d) Measured kinetics of ring expansion for LBS 4 [white arrow in (a,b)] for $V_g=1.22$ and 1.245 V at $P_{ex}=0.76$ mW (circles, triangles) and simulated kinetics for $\gamma=3$ (lower colorscale curve) and $4.95 \times 10^6$ s$^{-1}$ (upper colorscale
curve), $n_{bc}=12600$, $n_{bo}=0.05$. $D_e=80$ cm$^2$/s, $D_h=26$ cm$^2$/s, and $w=64$ in the simulations in (c,d).(e) Simulated kinetics for $D_e=30, 80,$ and 200 cm$^2$/s and experimental kinetics for $V_g=1.22$ V with the rest of the parameters the same as in (d). The electron diffusion constant $D_e=80$ cm$^2$/s provides the best fit to the measured kinetics.} \label{3}
\vspace{-4mm}
\end{figure*}

The parameters in simulations were chosen to fit the ring radius $R$ at the laser pulse end and the ring collapse time $\tau_R$, evaluated as the time it takes for the radius to drop by $50\%$. The simulations show that $R$ and $\tau_R$ depend strongly on some of the parameters, e.g. the hole and electron sources in Eq.(\ref{eq1}), and weakly on the others, e.g. the exciton formation rate $w$ or electron diffusion coefficient $D_e$ (APPENDIX \ref{app:fitting}).

Our task was made simpler because the radius $R$ is essentially independent of the hole diffusion coefficient (Fig. 2d). This can be understood from the continuity equation for hole transport in a steady state: for large $R$, the hole flux at the ring becomes independent of $D_h$ since an increase in $D_h$ leads to a corresponding reduction in the density gradient conserving the total hole flux \cite{typo}. In contrast, the collapse time $\tau_R$ depends strongly on $D_h$ and relatively weakly on $D_e$ (Fig.2e), since the electron source is spread everywhere in the CQW plane, while holes are only created in the laser spot and have to travel to the ring from there. These observations can be used to estimate $D_h$ from the observed ring kinetics: First, the sources can be determined by fitting the ring radius, then $D_h$ becomes the main parameter determining $\tau_R$, so that it can be estimated from fitting the measured $\tau_R$. The best fit gives an estimate $D_h = 26$ cm$^2$/s (Fig. 2c,e).

Now we analyze the kinetics of LBS rings (Fig. 3), which are centered around electron current filaments \cite{Butov04}. Each such filament represents a local electron source creating an electron-rich region, surrounded by a hole-rich region inside the external ring. The interface between the hole-rich and electron-rich regions is seen as an LBS ring in exciton emission (see schematic in Fig. 1j). The LBS-ring kinetics after the laser pulse was modeled by Eq.(1) with electron source peaked at the LBS center: $n_b(r \le r_0)=n_{bc},\quad n_{b}(r > r_0)=n_{bo}$. We used $r_0=0.26 \mu$m, which is smaller than the experimental resolution. The results were not sensitive to the choice of $r_0$.

When the laser is turned off, the hole source is terminated and the electrons injected by the current filament travel outwards expanding an electron-rich region around the LBS center and pushing the electron-hole interface away from it. This results in the expansion of the LBS ring (Fig. 3). Figure 3a-c shows that a brighter LBS expands faster. Since the LBS 1, 2 and 3 are at the same distance from the laser spot and thus have a similar hole density around them, this behavior originates from a larger electron source for a brighter LBS. We observe faster LBS ring expansion for higher gate voltage $V_g$; this originates from a larger electron source for a higher $V_g$. These dependences are consistent with those found in simulations. Since for LBS formation the roles of electrons and holes interchange, the LBS expansion and collapse kinetics can be used to determine the electron diffusion coefficient $D_e$ in a manner similar to the external ring kinetics used to determine $D_h$. The local electron source at the LBS center, which dominates the initial expansion of the LBS ring, can be determined from measured initial expansion speed. Following the steps outlined above we arrive at $D_e = 80$ cm$^2$/s.

In summary, our real-time imaging of exciton ring kinetics provide new insight into the physical processes underlying pattern formation in this system. In particular, exciton rings preserve their integrity during expansion and collapse, indicating that the kinetics is controlled solely
by in-plane carrier transport, which is a considerably slower process than exciton formation and recombination at the electron-hole interface. This property of ring kinetics allows us to perform a contactless measurement of carrier transport characteristics of the system, such as electron and hole diffusion coefficients.

This work is supported by ARO and NSF.

\begin{appendix}

\begin{figure*}\includegraphics[width=0.95\textwidth]{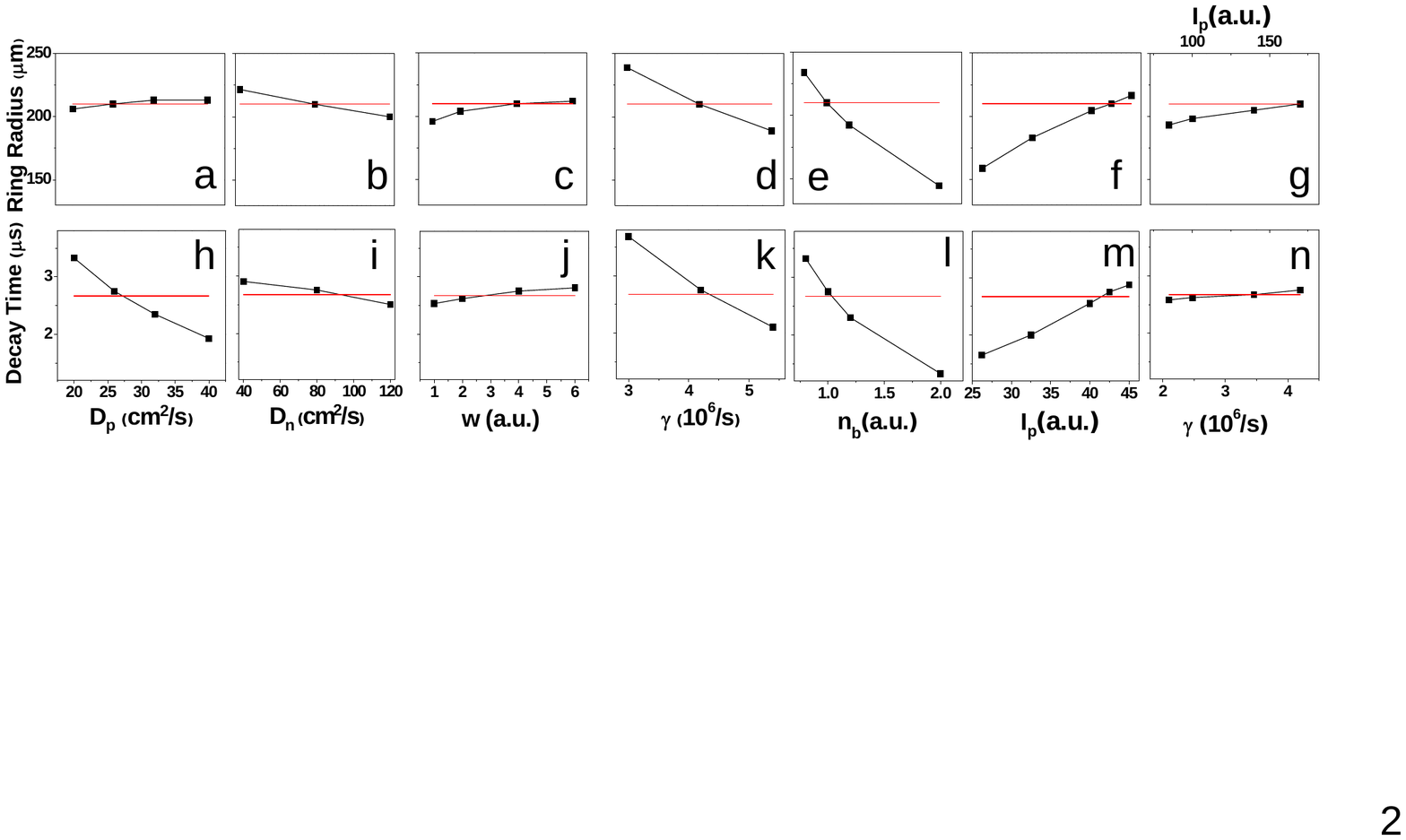}
\caption{(a-g) Radius and (h-n) collapse time of the external ring as a function of parameters in the simulations. (a-f, h-m) One parameter is varied while the others are kept at the values presented for the top curve in Fig. 2(a). $D_p$ and $D_n$ are the hole and electron diffusion coefficients, $w$ electron-hole binding rate, $\gamma$ electron escape rate, $n_b$ background electron density in the CQW in the absence of photoexcitation, and $I_p$ hole production rate in the excitation spot. (g,m) $I_p$ and $\gamma$ are varied simultaneously keeping the ratio $I_p/\gamma$ constant while the other parameters are kept at the values presented for the top curve in Fig. 2(a).} \label{a1}
\end{figure*}

\section{Accuracy of the model.}
 \label{app:accuracy}
 The approximations used in our model: (i) The drift of electrons and holes originating from the screened Coulomb interaction is neglected. This is accurate when $n, p \lesssim k_BT \varepsilon / (4 \pi e^2 L$), where $\varepsilon$ the dielectric constant, $L$ width of the intrinsic layers around the CQW; (ii) The hole escape out of the CQW is neglected. This is accurate when the number of escaping holes is small compared to the number of holes recombining with electrons in the CQW. Including such term in Refs. [16,17] of the main text results in a similar pattern; (iii) The hole source width is approximated as zero. This is accurate when it is much smaller than the ring diameter; (iv) The diffusion coefficients are approximated as coordinate and density independent; (v) The electron source is approximated as coordinate independent. The electron source is strongly enhanced at LBS centers in the CQW as described in the main text. This approximation is accurate when the ring area contains many LBS.

\vskip5mm

\section{ Unit calibration.}
\label{app:unit}
The model includes the units of length, time, and particle number. The units in simulations $L,T,$ and $N$ correspond to the real units in experiments $L', T',$ and $N'$ with the re-scaling factors
$C_L=\frac{L'}{L}, C_T=\frac{T'}{T}, C_N=\frac{N'}{N}$.

Experimental data for $P_{ex} = 1.82$ mW, which produces the ring with the radius $R' = 200$ $\mu$m in the experiment, correspond to the simulation with $I_p = 42.5$ and $R = 8$, see Fig. 2a in the main text. This yields $C_L=\frac{R'}{R} = 25\mu m$. The laser pulse width in the experiments $T' = 10 \mu$s corresponds to $T = 300$ in the simulations. Therefore $C_T=\frac{T'}{T} = 3.3 \times 10^{-2}\mu$s. The determination of $C_N$ requires the measurement of either the formation rate of excitons or the current through the structure, see below. It is not done in the present work.

In the equation used in the simulations
\begin{equation}\label{main}
   \begin{array}{l}
   \partial_{t} n = D_n\nabla^2n-wnp+\gamma(n_b-n),  \\
   \partial_{t} p = D_p\nabla^2p-wnp+I_{p}\delta(\vec{r}),  \\
   \end{array}
\end{equation}
all parameters are in the simulation units. The equation should keep its form when the units are changed to real. The rescaling relationships for the terms in the equation are: $\partial_{t} = C_T \partial_{t'}, \nabla^2=C_L^2\nabla'^2, n=\frac{C_L^2}{C_N}n', I=\frac{C_T}{C_N}I', \delta (\vec{r})=C_L^2\delta (\vec{r'})$. Plugging them to Eq. (1) yields:
\begin{equation}
   \begin{array}{l}
   \frac{C_TC_L^2}{C_N}\partial_{t'} n' = \frac{C_L^4}{C_N}D_n\nabla'^2n'-\frac{C_L^4}{C_N^2}wn'p'+\frac{C_L^2}{C_N}\gamma(n'_b-n'),  \\
   \frac{C_TC_L^2}{C_N}\partial_{t'} p' = \frac{C_L^4}{C_N}D_p\nabla'^2p'-\frac{C_L^4}{C_N^2}wn'p'+\frac{C_TC_L^2}{C_N}I'_{p}\delta(\vec{r'}).  \\
   \end{array}
\end{equation}
Comparing Eq. (2) with with the equation in real units
\begin{equation}
   \begin{array}{l}
   \partial_{t'} n'=D_n'\nabla'^2n'-w'n'p'+\gamma'(n_b'-n'),  \\
   \partial_{t'} p'=D_p'\nabla'^2p'-w'n'p'+I'_{p}\delta(\vec{r'}),  \\
   \end{array}
\end{equation}
gives the rescaling relationships for the coefficients:
\begin{equation}
   \begin{array}{l}
   D_n'=\frac{C_L^2}{C_T}D_n=200cm^2/s\times D_n,  \\
   D_p'=\frac{C_L^2}{C_T}D_p=200cm^2/s\times D_p,  \\
   w'=\frac{C_L^2}{C_TC_N}w=\frac{200cm^2/s}{C_N}\times w, \\
   \gamma'=\frac{1}{C_T}\gamma=3\times10^7/s\times\gamma.
   \end{array}
\end{equation}

As described in the main text, $D_n$ can be obtained from the LBS ring kinetics and $D_p$ -- from the external ring kinetics. The other parameters are briefly discussed below. The parameter describing the electron source $\gamma$ can be obtained from both LBS and external ring measurements as described in the main text. Our simulations show that varying the exciton formation rate $w$ affects the ring width but practically does not change the ring radius and decay time. Therefore, $w$ was not determined from the data. The determination of the density $n = \frac{C_L^2}{C_N}n'$ requires the evaluation of $C_N$. In turn, $C_N$ can be determined by measuring the electron current through the sample in the absence of photoexcitation. However, the current in the experiments is small and its measurement is beyond the scope of this work (total leakage current in the structure is $< 1 \mu$A for the excitations above the AlGaAs gap and below the instrumentation sensitivity 10 nA for resonant or no excitation).

\vskip5mm

\section{Fitting procedure}.
\label{app:fitting}
There are six parameters in Eq. \ref{main}:
diffusion coefficients $D_n$ and $D_p$; exciton formation rate $w$; and source terms presented by $I_p$, $\gamma$, and $n_b$. For $C_N$ unknown, we set $n_b = 1$. We probed the dependence of the ring radius and collapse time on each other parameter, see Fig. \ref{a1}.

Varying the exciton formation rate $w$ practically does not change the ring radius $R$ and collapse time $\tau_R$, see Fig. \ref{a1}c,j. Varying $I_p$ and $\gamma$ simultaneously keeping $I_p/(\gamma n_b)$ constant only weakly changes $R$ (consistent with the conclusions of Ref. [17]) and $\tau_R$, see Fig. \ref{a1}g,n.

In our fitting procedure for the external ring, we first determined the sources by fitting $R$, see Fig. \ref{a1}d-f. Then $D_h$ becomes the parameter, which determines the ring collapse time, so that it can be estimated from fitting the measured $\tau_R$. The best $\tau_R$ gives an estimate $D_h = 26$ cm$^2$/s, see Fig. \ref{a1}h. Both $R$ and $\tau_R$ are only weakly sensitive to $D_n$, see Fig. \ref{a1}b,i. The estimate for $D_n = 80$ cm$^2$/s was obtained by fitting the kinetics of LBS rings, see the main text. In turn, the obtained values of $D_n$ and $D_p$ were used in estimating $\gamma, n_b$ and $I_p$ by fitting the ring radius, Fig. \ref{a1}d-f, embracing the procedure.

The procedure for LBS was similar. The ratio of sources was obtained by fitting the initial LBS ring expansion between $t=0$ and $1 \mu$s. Then $D_n$ was estimated by fitting the expansion at $t > 1 \mu$s.

The false color plots presenting the simulations in the paper show the value $n \times p$, which is proportional to the exciton density. To combine several simulation results in one figure, we present the maximum value in the simulations for each time.

\end{appendix}


\begin{thebibliography}{99}

\vspace{-4mm}

\bibitem{Smith89}
L.M. Smith, J.S. Preston, J.P. Wolfe, D.R. Wake, J. Klem, T. Henderson, H. Morkoc,
Phys. Rev. B {\bf 39}, 1862 (1989).

\bibitem{Sivan92}
U. Sivan, P.M. Solomon, H. Shtrikman, Phys. Rev. Lett. {\bf 68},
1196 (1992).

\bibitem{Eisenstein92}
J.P. Eisenstein, G.S. Boebinger, L.N. Pfeiffer, K.W. West, S. He,
Phys. Rev. Lett. {\bf 68}, 1383 (1992).

\bibitem{Lay94}
T.S. Lay, Y.W. Suen, H.C. Manoharan, X. Ying, M.B. Santos, M. Shayegan,
Phys. Rev. B {\bf 50}, 17725 (1994).

\bibitem{MacDonald01}
A.H. MacDonald, Physica B {\bf 298}, 129 (2001).

\bibitem{Pellegrini07}
V. Pellegrini, S. Luin, B. Karmakar, A. Pinczuk, B.S. Dennis, L.N. Pfeiffer, and K.W. West,
J. Appl. Phys. {\bf 101}, 081718 (2007).

\bibitem{Tiemann08}
L. Tiemann, J.G.S. Lok, W. Dietsche, K. von Klitzing, K. Muraki, D. Schuh, W. Wegscheider,
Phys. Rev. B {\bf 77}, 033306 (2008).

\bibitem{Croxall08}
A.F. Croxall, K. Das Gupta, C.A. Nicoll, M. Thangaraj, H.E. Beere, I. Farrer, D.A. Ritchie, M. Pepper,
Phys. Rev. Lett. {\bf 101}, 246801 (2008).

\bibitem{Seamons09}
J.A. Seamons, C.P. Morath, J.L. Reno, M.P. Lilly,
Phys. Rev. Lett. {\bf 102}, 026804 (2009).

\bibitem{Keldysh65}
L.V. Keldysh, Yu.E. Kopaev,
Fiz. Tverd. Tela {\bf 6}, 2791 (1964) [Sov. Phys.
Solid State {\bf 6}, 6219 (1965).

\bibitem{Keldysh68}
L.V. Keldysh, A.N. Kozlov,
Zh. Eksp. Teor. Fiz. {\bf 54}, 978 (1968) [Sov. Phys.
JETP {\bf 27}, 521 (1968).

\bibitem{Butov02}
L.V. Butov, A.C. Gossard, D.S. Chemla, Nature {\bf 418}, 751 (2002).

\bibitem{Ivanov06}
A.L. Ivanov, L.E. Smallwood, A.T. Hammack, Sen Yang, L.V. Butov, A.C. Gossard,
Europhys. Lett. {\bf 73}, 920 (2006).

\bibitem{compare}
In our experiment, transport of electrons or holes in one layer is probed essentially in the absence of carriers in the other layer.

\bibitem{Butov04}
L.V. Butov, L.S. Levitov, A.V. Mintsev, B.D. Simons, A.C. Gossard, D.S. Chemla,
Phys. Rev. Lett. {\bf 92}, 117404 (2004).

\bibitem{Rapaport04}
R. Rapaport, G. Chen, D. Snoke, S.H. Simon, L. Pfeiffer, K. West, Y. Liu, S. Denev,
Phys. Rev. Lett. {\bf 92}, 117405 (2004).

\bibitem{Haque06}
M. Haque, Phys. Rev. E {\bf 73}, 066207 (2006).

\bibitem{Chen05}
G. Chen, R. Rapaport, S.H. Simon, L. Pfeiffer, K. West,
Phys. Rev. B {\bf 71}, 041301 (2005).

\bibitem{Hammack07}
A.T. Hammack, L.V. Butov, L. Mouchliadis, A.L. Ivanov, A.C. Gossard,
Phys. Rev. B {\bf 76}, 193308 (2007).

\bibitem{typo}
Formula (3) in \cite{Butov04} for the ring radius contains a typo, which has been corrected in Eq. (8) in \cite{Haque06}.

\end{thebibliography}
\end{document}